\documentclass[aps,amsmath,prb,a4paper,showpacs,twocolumn]{revtex4}
\usepackage{graphicx}

\newcommand{\mcal}{{\cal{M}}}
\newcommand{\beq}{\begin{equation}}
\newcommand{\eneq}{\end{equation}}
\newcommand{\bea}{\begin{eqnarray}}
\newcommand{\enea}{\end{eqnarray}}
\newcommand{\bean}{\begin{eqnarray*}}
\newcommand{\eean}{\end{eqnarray*}}

\begin{document}
     
\title{Superconductive proximity in a Topological Insulator slab and excitations bound to an axial vortex}

\author{A. Tagliacozzo$^{1,2}$} \author{P. Lucignano $^{3,1}$}  \author{F.Tafuri $^{4,2}$} 

\affiliation{$^1$ Dipartimento di Scienze Fisiche, Universit\`a di Napoli Federico II, Monte S.Angelo via Cinthia 80126 Napoli Italy} 
\affiliation{$^2$ CNR-SPIN, Monte S. Angelo via Cinthia 80126 Napoli Italy} 
\affiliation{$^3$ CNR-ISC, sede di Tor Vergata, Via del Fosso del Cavaliere 100, 00133 Roma, Italy} 
\affiliation{$^4$ Dipartimento di Ingegneria, Seconda  Universit\`a di Napoli, Aversa, Napoli Italy} 

\begin{abstract}
We consider the proximity effect in a Topological Insulator sandwiched between two conventional superconductors, by comparing $s-wave$ spin singlet superconducting pairing correlations and $odd-parity$ triplet pairing correlations with zero spin component orthogonal to the slab ("polar " phase). A superconducting gap opens in the Dirac dispersion of the surface states  existing at the interfaces. An axial vortex  is included, piercing the slab along the normal to the interfaces with the superconductors. It is known that, when proximity is  $s-wave$,  quasiparticles in the gap are Majorana Bound States, localized at opposite interfaces. 
We report the full expression for the quantum field associated to the midgap neutral fermions, as derived  in the two-orbital band model for the TI.   When proximity involves  $odd-parity$ pairing,  midgap modes are charged Surface Andreev Bound States, and they originate from interfacial circular states  of  definite chirality, centered at the vortex singularity and decaying in the  TI film with oscillations.   When the chemical potential is moved away from midgap, extended states along the vortex axis are also allowed. Their  orbital structure depends on the symmetry of the bulk band from where the quasiparticle level  splits off. 
    \end{abstract}
\pacs{73.20.-r, 74.20.Rp, 74.45.+c}

\maketitle

\section{Introduction}
Topological insulators (TIs), bulk  insulators, with metallic surface states protected by time reversal invariance, are attracting  widespread interest for potential applications in nanoelectronics and spintronics \cite{hasankane,qizhangmod}.
Three-dimensional (3D) TIs, such as $Bi_2Se_3 $ or $ Bi_2Te_3 $ \cite{hasan, zhang_nphys}, have two dimensional (2D) metallic surface states, whose band structure consists of an odd number of Dirac cones, centered at  Time Reversal   (TR) invariant  momenta in the surface Brillouin Zone (BZ). 
Surface sensitive experiments such as angle-resolved photoemission spectroscopy  \cite{hasan3, nishide} and scanning tunneling microscopy (STM)\cite{wang} have confirmed the existence of these exotic surface metal states with a Dirac-like  energy dispersion.  

Boundary states can be characterized with the help of  minimal models for the various geometries, whose corresponding Hamiltonians are classified generalizing  Altland-Zirnbauer symmetry classes \cite{altlandzirnbauer} and  the  topological invariants in  mapping  from the configurational space, parametrized by the $k-$ vectors,  to the real  space.
Surface Dirac fermions can become chiral when time reversal is broken by (e.g.) ferromagnetic covering  and they can be bound to defects such as magnetic domain walls. The zoology of all the possible boundary states available has been presented in full in Ref.s \onlinecite{teokane, schnider}. 
 A model which successfully describes e.g. the (1,1,1) surface of the (otherwise centrosymmetric) $Bi_2Se_3$  close to the $\Gamma $ point of the surface BZ  has been introduced in Ref.  \onlinecite{zhang_nphys} and can be taken as the prototype model for  TR invariant  TIs. It is based on  the continuum limit for long wavelengths and includes appropriately the spin-orbit interaction which is the crucial feature for the topological protection. It is a two band model involving two species of orbitals, of even and odd parity, with a bulk gap $2 M$. 
 
  We consider a   TI film terminated at an $ x-y $ plane,  sandwiched between two superconductors.  Here we choose  $\hat z $  perpendicular to the quintuple layers \cite{fanzhang}. The metallic states at the interfaces become superconducting by proximity effect. The nature of the induced superconductivity when the chemical potential is within the bulk  gap is unknown. Solving this  puzzle is extremely important in view of the possible existence of   Majorana Bound States (MBSs)  localized at interfaces between conventional superconductors and TIs\cite{kitaev}. It has been argued that  induced singlet, $s-wave$  superconducting correlations,  with order parameter $\Delta _s $, could turn the pairing of the boundary states  (with $ M> \mu >> \Delta _s $) into an effective spinless 
 $odd-parity$ pairing of the $p-wave$ type: $p_x -i p_y$\cite{fucane100}.  In this case a vortex piercing the structure can bind a Majorana fermion at each interface\cite{fucane100,kopnin,readgreen,ioselevich}. It has been shown that,  for the case of an $s-wave$ pairing, in presence of an axial vortex,  the zero energy Majorana Fermion can survive even when $\mu $ is moved within one of the bulk bands by p- or n-doping, till the spin-orbit coupling is able to supply a Berry phase of  $\pi $ to its wavefunction\cite{hosur}. Roughly speaking, the minus sign that the Majorana fermion acquires in turning once around a loop, which is intrinsic to its fermionic character,  should be compensated by the spin-orbit texture, in order to provide a single valued solution of the eigenvalue problem.

Still, there is no evidence that superconductivity induced in the boundary states would be of the singlet, $s-wave$ type.

   It has been found that  few per cent  $ Cu $ doping  of   $Bi_2Se_3$ bulk crystals moves the chemical potential $\mu$ within the bulk conduction band and  the superconducting phase transition takes place at $T_c \sim 3.8K $\cite{hor,wang}.   Odd-parity pairing  seems to be favored by strong spin-orbit coupling\cite{fuberg}. Topological odd-parity superconductors are fully gapped in the bulk but have gapless Surface Andreev Bound States (SABS). It has been noticed recently  that a mirror symmetry (composed of  the odd mirror reflection in the $yz$ plane $x \to -x$ and  of a $Z_2 $ gauge transformation  $\Delta \to -\Delta $) characterizes the helicity of the  SABS\cite{qizhangmod,hsiehfu}.
Superconductivity induced in the boundary states of  a  TI  slab  by a singlet superconductor,  odd with respect to the mirror plane $z=0$ could be of the "polar" type, as explained below, with an order parameter proportional to $p_z$ and a vanishing spin projection onto the $z$ spin quantization axis, which is pinned  to the normal to the slab by the spin-orbit interaction. Strictly speaking, such a superconducting order would have a nodal line of excitations  exactly at the $x-y$  plane, which allows for the sign inversion at the boundary. However  state of the art  TI flakes  are known to be plagued by impurities and vacancies in the bulk and it is likely that  bulk  impurity  bands  arise with little dispersion which provide a non vanishing Fermi momentum $p_F$ also in the $z-$direction. Therefore, we argue that the proximity gap never closes, even when $ \mu $ is located within the bulk gap. Otherwise, the presence of the  gap nodal line at the surface would  affect the lifetime of  the states bound at a vortex core, but not their topological  origin.  

In this paper we explore the nature of  the zero energy excitations corresponding to quasiparticles bound  to an axial  vortex piercing  a sandwich  $S/TI/S $ geometry and carrying a magnetic flux. While  topological superconductors  can be well classified when TR invariance is preserved\cite{qihuges, zhang2},  by the breaking of TR symmetry the topological protection can be washed out.   We compare the two kind of superconducting induced orders  in the limit $ M> \mu >> \Delta $:  $a)$  an s-wave  induced order parameter, $b)$   a $p_z$-wave (polar), odd parity order parameter. 

The two-orbital model appears to be adequate for  both the cases. Using this model we find that a  proximity superconducting gap opens at $ \mu $  located in the Dirac cone dispersion of the boundary states. A vortex piercing the slab binds zero energy excitations inside the superconducting gap depending on the value of $ \mu $. However, their nature is strongly dependent on the type of superconducting ordering, $a)$ or $b)$,  induced by proximity:
\begin{description}
\item[a)]   Majorana  zero energy excitations are bound to the vortex as expected\cite{fucane100} (called MBS henceforth).  A   neutral quasiparticle  is localized  at each of the interfaces with the superconductors and we present an analytical explicit  expression for the quantum field that describes the excitation in the two-orbital model in the simple case of $\mu = 0 $ (midgap state).  Orbital parities are mixed and spin is not conserved. The two spacially separated modes are TR mates, no matter that the vortex breaks TR symmetry. It has been argued that even moving the chemical potential inside the bulk band the vortex could host a MBS\cite{hosur}.
\item[b)]  In the case of odd parity topological superconductivity, with $\mu$  inside the bulk bands,  a linear dispersion of  SABS  arises  in the superconducting gap, close to $ k=0$\cite{hsiehfu}. These states are MBS.   In this work we show that when $\mu $ is in the bulk gap, a vortex hosts SABS, as well. However they are no longer neutral fermion excitations. By  breaking  the TR symmetry of the odd-parity topological superconductor,  the  vortex  makes the  zero energy SABS  excitations turn into a pair of  Dirac  states, loosely bound to the vortex core.  To prove our statement, we solve explicitely the $\mu =0$(midgap) case, in full analogy with the $s-wave$ pairing case described in  $a)$. The modes,  having odd  and even  orbital symmetry, respectively,  propagate  along  the vortex axis with opposite chirality. At $\mu \neq 0$,   if the chemical potential  matches an energy split off from the bulk conduction or valence  band, zero energy excitations exist, with mixed type of orbitals.  Expectation value of the spin of  these excitations is, in any case, zero. 
 \end{description}
 The important novelty in our calculation is the assumption of  polar order parameter for $\mu $ within the bulk gap in the two-orbital model.  Polar order, i.e. triplet pairing with zero projection of the Cooper pair spin along the normal to the boundary plane, is expected to be preferred in view of strong tendency to in-plane helical transport induced by spin-orbit coupling .  We fully account for the two bulk band structure in the TI and for the  third dimension $\hat {z}$ orthogonal to the plane of the slab. The  orthogonal direction is crucial for the description of these bound states. The celebrated argument by Fu and Kane \cite{fucane100} is based on an effective two-dimensional system describing the flat boundary.  It is easy to see that its generalization  is unable to account for the proximity in the polar state. This is what we shortly report  on, in closing this Introduction. 
 
   The  starting point of  Fu and Kane \cite{fucane100}  is the  Dirac dispersion of the surface electronic states,  proximized by an $s-wave$ superconductor. The model  can be effectively  reduced to  spinless electrons with linear energy dispersion in a $p-wave$ pairing field. Here we want to extend their argument to induced odd parity  proximization.   

The original argument can be rephrased as follows. We consider just an effective Dirac-like Hamiltonian for the surface states at the flat boundary at $z=0$.The velocity $v$ characterizes the linear dispersion.   The  proximized TI Hamiltonian,
in the basis $ [ \psi _{k\uparrow}, \psi _{k\downarrow}, \psi _{-k\downarrow} ^\dagger, - \psi _{-k\uparrow} ^\dagger ] $, is\cite{not}:
\beq
 H_{TI} =  \left (\begin{array}{cc}  \begin{array}{cc} -\mu &v k_- \\ vk_+ &-\mu\end{array} & \hat{\Delta} \\ \hat{ \Delta} & \begin{array}{cc} \mu & -vk_- \\  -vk_+ & \mu \end{array}\end{array}\right ) ,
\eneq
with $k_\pm = k\: e^{\pm i \: \theta _{\vec{k}} } $.  In this basis, $s-wave$ pairing takes the form $ \hat{\Delta}_s = \left (\begin{array}{cc} \Delta _s& 0 \\ 0 &\Delta _s\end{array} \right )$ and  the  four energy eigenvalues are $ \pm \sqrt{\left( \pm vk -\mu \right )^2 + \Delta _s ^2 }Ê$.
 If we disregard the two bands which are far off the chemical potential,   it is easy to show that an effective BdG Hamiltonian can be written in terms of spinless operators corresponding to  spinorial wavefunctions: 
   \begin{widetext}
   \beq
c_{\vec{k}}  \equiv  \left [\begin{array}{c} \psi_{\vec{k} \uparrow }\\  \psi_{\vec{k} \downarrow } \end{array} \right ]\: \to
\frac{1}{\sqrt{2}} \: \left [\begin{array}{c} 1\\ e^{i\theta _k}  \end{array} \right ]\: ; \:\:\:\:   
  c_{-\vec{k}}\:  ^\dagger  \equiv  \left [\begin{array}{c} \psi_{-\vec{k} \downarrow } \: ^\dagger \\  - \psi_{-\vec{k} \uparrow}\:  ^\dagger \end{array} \right ]\: \to  \frac{1}{\sqrt{2}} \: 
\left [\begin{array}{c} e^{-i\theta _{-k}}  \\ -1\end{array} \right ] \: .
\eneq
The new effective Nambu Hamiltonian  reads:
\beq
   H^{new}   = \frac{1}{2} \: \begin{array}{c} \left ( c_{\vec{k}} \: ^\dagger   \:\:   - c_{- \vec{k}}     \right )  \\  . \end{array}  \:    \left ( \begin{array}{cc}  \left [vk -\mu \right ]  & \Delta_s e^{-i\theta _k}   \\  \Delta_s  e^{ i\theta _k} & - \left [vk -\mu \right ]   \end{array}\right )  \left (  \begin{array}{cc}  c_{\vec{k}}  \\ -  c_{-\vec{ k}} \: ^\dagger  \end{array} \right )  
   \label{fuit}
   \eneq
   \end{widetext}
   Indeed, e.g., the matrix element ,  $ h_{12}^{new} , $ in the basis given above is:
\bea
     h^{new} _{12}  = \frac{1}{2} \: \begin{array}{c} \left ( 1  \:\:   e^{-i\theta _k}   \right )  \\ \end{array}  \:    \left ( \begin{array}{cc} \Delta _s & 0  \\ 0 & \Delta _s  \end{array}\right )  \left (  \begin{array}{cc}  e^{-i\theta _k}  
   \\  1\end{array}  \right )  =  \Delta_s \:  e^{-i\theta _k} .  \nonumber
   \enea
 Eq.\eqref{fuit}  provides  an effective   Hamiltonian  for spinless  particles with $\Delta _{\vec{k} } \propto k_x-ik_y $, which implies an effective $p-wave$ pairing. According to ref.\onlinecite{kopnin},\onlinecite{readgreen}, a vortex can sustain a MBS in such a  system. 
 
 On the other hand, the odd-parity mean field coupling   $\sum _{\sigma  \sigma '}  \Delta _{\sigma ' \sigma } \:   \psi _{\vec{k}\sigma } ^\dagger  \psi _{-\vec{k} \sigma ' }^\dagger  \:  +  h.c. $   is described by a vector $ \vec{d} (\vec{k}) $ which is   odd for $  \vec{k} \to -  \vec{k} $. The generic, uniform  order parameter is a  matrix  in the spin space  
$\Delta _{\sigma \sigma'} \left ( \vec{k} \right ) = \left [  \vec{d} \left ( \vec{k} \right ) \cdot  \left ( \vec{s} \;  i s_2\right ) \right ]_{\sigma\sigma'} $.  In our basis, it reads:  $ \hat{\Delta}_p = \left (\begin{array}{cc} d_z& d_x-id_y \\ d_x+id_y &-d_z\end{array} \right )$.   If $ \vec{d}  ( \vec{k}  ) \propto  \vec{k} $ itself, the effective Hamiltonian $H^{new} $ is the same as in eq.\eqref{fuit}, but with $\Delta _s $ replaced by $ |d_\parallel | $ with  $d_\parallel \equiv \left ( d_x,d_y  \right )$. In the model  $d_z$ disappears altogether.
 This result doesn't look convincing, because it seems to be  an artifact due to the restriction of the model to the   two-dimensional  $x-y$ boundary plane. By contrast, in view of the strong spin-orbit locking between spin and momentum in this plane, it is likely  that a polar "p-wave" phase is stabilized, with vanishing $z-$projection of the Cooper pair spin. This would imply an order parameter  $  \vec{d}  \equiv \left  (0,0,d_z \right ) \propto   k_z \hat {z} $. In the bulk, such an order parameter would have a   gap with a nodal line  at the Fermi surface, exactly at $k_z =0 $, because  $Det (\hat{\Delta} _p)   \propto k_z ^4$.  However, this may not be the case close to the surface where $k_z$ is undefined. Besides,   state of art flakes of topological insulators (prepared, for instance, by Chemical Vapor Deposition) are not impurity free in the bulk, and we argue that a non vanishing Fermi wavevector in the $z-$ direction is anyhow present in the bulk,  thus providing a non vanishing superconducting gap also at the boundary surface.  Given this choice for the order parameter, our heuristic approach leading to Eq.\eqref{fuit} would make,  in any case, the superconducting order parameter disappear in $H^{new}$, what is suspicious.  
 
 This argument shows that, when odd parity symmetry for the superconducting order parameter is chosen, with vanishing 
 spin component of the pair in the direction orthogonal to the surface, a full approach which includes the third dimension is  mandatory.

The structure of the paper is the  following.
 In Sec. II we  first introduce the two-orbital model  in the long wavelength continuum limit, close to the $\Gamma $ point, which is of the Bogolubov-De Gennes type \cite{degennes}. The model applies to a bulk crystal with superconducting pairing correlations. It will be convenient  to deal with  the $odd-parity$ order parameter case first, and rephrase the model  for the $s-wave$  pairing. This is done  in Sec. III, where  we derive the opening of the superconducting gap at the chemical potential, which can be located within one of the two bulk bands (thus giving rise to what is called a topological superconductor, Eq.\eqref{bubu}), or within the bulk gap. In the latter  case the superconducting gap opens in the  energy dispersion of the  boundary states which is Dirac cone like, Eq.\eqref{enercone}. The neutral excitation  bound to the  vortex at the slab boundary  in the case of $s-wave$ proximity is derived analytically for   in Appendix A and discussed in  Sec. IV.A. In Sec. IV.B we report on the midgap ($\mu =0 $) charged  excitations bound at the vortex core in the  $odd-parity$  pairing. Their field operator  and its spacial dependance is  derived in  Appendix B.  Extended  states along the vortex core are also derived, for the case of $\mu \neq 0$.  Results are summarized in Sec. V. 
 From our analysis, the main  conclusion that we draw is  the observation that  there are no MBS with   $odd-parity$  superconductive proximity and zero projection of the pair spin, in the two-orbital model.

\section{Model Hamiltonians} 
We start from the Hamiltonian introduced by H. Zhang $ et \:  al. $ \cite{zhang_nphys} to describe  three dimensional  layered systems like $Bi_2Se_3$ $Sb_2Te_3$, and $Bi_2Te_3$. They use a $4$-dimensional  basis-space which is the direct product of the  spin space and of the orbital parity space originating from  $p_z$ orbitals, denoted as:
$\left \{Ê|p_z^g, \uparrow \rangle , Ê|p_z^u, \uparrow \rangle , Ê|p_z^g, \downarrow \rangle , 
Ê|p_z^u, \downarrow \rangle  \right \} $. Here  $g(u) $ denotes even (odd) parity for  $\vec{k} \to -\vec{ k} $. On such  a basis, the model Hamiltonian is (we assume   the Fermi velocity $v$ to be isotropic and we put it equal to unity  here and in the following):
\bea
\hat h_o  =  -{\cal{M}} \sigma _3  + i \sigma _1 \left ( s_1 \partial _x - s_2 \partial _y\right )  +i\sigma _3 s_3 \partial _z \: .
\label{ham0}
\enea
Here $s_\alpha$  and  $\sigma _a$ ($\alpha,a \in \{1,2,3\}$) are  Pauli  matrices in the spin and orbital space, respectively. The hat, over $\hat h_0$ reminds of its 4x4 matrix structure.  ${\cal{M}}$ includes second order derivatives ${\cal{M}} = M  (z)+C\: \partial _{\parallel}^2 +c \:\partial ^2_z  $, parallel to the flat  $x-y$ boundary plane terminating  the quintuple layer.    $ M $ is  half the bulk gap.
Non trivial topology is guaranteed by the condition $M,C,c >0 $, which corresponds to the  inversion of the bulk bands. 

The Bogolubov-De Gennes (BdG) mean field Hamiltonian, in the presence of a superconducting pairing requires an 8 x 8 matrix structure whose compact form is:
\beq
H_{BdG} (\vec{k})  =   \left (  \begin{array}{cc} 
 \hat{h}_o&   \hat\Delta  \\
\hat{ \Delta} ^\dagger & - \hat{h}_o^*
\end{array} \right )  \: ,
\label{mat2} 
\eneq
where  $  \hat{h}_o $ and  $   \hat\Delta $  are  $4X4 $ matrices. In the BdG representation, even parity singlet pairing requires that  $\hat  \Delta _s ^T  (\vec{k}) = \hat{\Delta} _s (-\vec{k})  $, while odd-parity  triplet pairing satisfies  $\hat  \Delta _p ^T  (\vec{k}) = -\hat{\Delta} _p (-\vec{k})  $\cite{ivanovbook}.

In our case, given the Hamiltonian 
\beq
H -\mu \:  {N}   =  (  h_0 -\mu)\tau_z + H_{pair}  
\label{bdg} 
\eneq
where the $\tau_a$ Pauli matrices act in the Nambu space,  the matrix structure of the off-diagonal pairing Hamiltonian $H_{pair}$ can acquire  different forms, depending on the order parameter symmetry.  In the even parity,  s-wave, singlet case, we define  $\Delta_s =   \langle \psi _{u\uparrow}  \psi _{u\downarrow} \rangle   =  \langle \psi _{g\uparrow}  \psi _{g\downarrow} \rangle  =-  \langle \psi _{u\downarrow}  \psi _{u\uparrow} \rangle = -\langle \psi _{g\downarrow}  \psi _{g\uparrow} \rangle $, assumed to be independent of the orbital.  It gives rise to a pairing Hamiltonian $$H_{pair}^s= -i (\Delta_s  s_y  \tau_+ + h.c.).$$

On the other hand,  it has been proposed that, when doped with few percent  $Cu$, the $ Bi_2Se_3 $ undergoes  the superconducting phase transition with an  odd-parity  order parameter \cite{fuberg}. 
  The pairing Hamiltonian for the polar ordering presented in the Introduction  is 
$$H^p_{pair} = (\Delta_p  \sigma_y s_y \tau_+ + h.c.),$$
where  ${ \Delta}_p =\langle \psi _{u\uparrow}  \psi _{g\downarrow} \rangle= - \langle \psi _{g\uparrow}  \psi _{u \downarrow} \rangle \:  =- \langle \psi _{g\downarrow}  \psi _{u\uparrow}  \rangle  =  \langle \psi _{u\downarrow}  \psi _{g\uparrow}  \rangle $  is the odd-parity orbital order parameter.  The change in signs in the expectation values for $\Delta_p $ arises from the fact that the operators act on a triplet pair with zero spin projection along $z$.  

\section{TI boundary  states}
In this Section  we discuss the opening of the superconducting gap induced by proximity in the TI, when  the bulk gap $M$ is much larger than the  superconducting gap $\Delta $. Depending on the doping, the  chemical potential $\mu$  can be within a bulk band with dispersion $\eta _{\vec{k}} \approx  \pm \sqrt{ M^2 + k^2 } $, or in the gap of the insulator, where  the electronic states localized at the boundaries  have a Dirac like dispersion.  Here we choose  $\hat z $  perpendicular to the quintuple layers \cite{fanzhang}, which would give helical boundary states in the surface plane. Spin orientation is expected to be in the plane, as well.  The Hamiltonian for the  two different pairing symmetries, which we have introduced  in the previous Section and the corresponding energy spectrum is given here below.  

\subsection*{Odd-parity  pairing}

  Let us first write down the Hamiltonian for  the odd-parity pairing.   It is convenient  for our purposes  to choose the basis 
\beq
 B\equiv \left [  \psi_{g\uparrow}, \psi_{u\downarrow},
 \psi_{u\downarrow}\;^\dagger ,- \psi_{g\uparrow }\;^\dagger | \psi_{u\uparrow}, \psi_{g\downarrow},
 \psi_{g\downarrow}\;^\dagger,  - \psi_{u\uparrow }\;^\dagger \right ]^T .
 \label{basic}
 \eneq
   A matrix acting on this basis has $ 4 X4 $  blocks corresponding to components with  exchanged parities ($g, u$). We address these blocks with  Pauli matrices denoted by  $T^a $. Within each of these blocks, a Nambu particle-hole component structure arises, of the kind: $ [ ( \psi _\uparrow,\psi _\downarrow),
 (\psi _\downarrow ^\dagger , - \psi _\uparrow ^\dagger )]^T $, addressed by Pauli matrices ${{\cal C}}^\alpha $. The spin  space  is addressed by Pauli matrices $S^\sigma $ ($ a,\alpha,\sigma =1,2,3)$).    In this representation, the    particle-hole  conjugation is defined by  $  \Xi  = \left ( {\bf 1} \times  {{\cal C}}^2\times   S^2 \right ) {\cal {K}} $, where  $ {\cal {K}} $ is the complex conjugation  of  the wavefunctions. 
 Explicitly, the  application  of $  \Xi  $ onto the basis $B$ gives,  apart for $ {\cal {K}}$:
 $\Xi B = 
  \left [ \psi_{g\uparrow}\: ^\dagger,  \psi_{u\downarrow}\: ^\dagger,
\psi_{u\downarrow} , - \psi_{g\uparrow }, \psi_{u\uparrow}\: ^\dagger,  \psi_{g\downarrow} \: ^\dagger , \psi_{g\downarrow},  - \psi_{u\uparrow }\right ]^T $.
  Time Reversal  operator is  $\Theta =  [ {\bf 1}  \times  {\bf 1}  \times  i S^2 ] \:{\cal{K}} $   ($\Theta ^2 = -1 $). 
  This  transforms the basis into
$ \left [  \psi_{u\downarrow} , - \psi_{g\uparrow} ,
- \psi_{g\uparrow}\;^\dagger , - \psi_{u\downarrow }\;^\dagger,   \psi_{g\downarrow} , - \psi_{u\uparrow}, - \psi_{u\uparrow}\;^\dagger , - \psi_{g\downarrow }\;^\dagger \right ] ^T $, apart for ${\cal{K} } $.

   The full $8X8 $  Hamiltonian in the basis  $B$  looks like:
 \beq
H^p \!\!=\!\! 
  \left  (  \:   \begin{array}{cc} H_+^p &   S^3\:  i\partial_z  \\
  S^3 \:   i\partial_z   & H_-^p 
\end{array} 
\right ),
\label{fullpo}
\eneq 
where  
 \beq 
 H_{\pm}^p \!\!\ = \!\! \left (  \begin{array}{cccc}  \mp \mcal-\mu & i \left (\partial_x -i \partial_y \right )& \pm \Delta_p  & 0\\
i   \left (\partial_x +i \partial_y \right ) & \pm \mcal -\mu& 0& \pm \Delta_p \\
 \pm \Delta_p ^* & 0 & \pm \mcal+\mu& i \left ( \partial _x -i\partial _y \right )  \\
 0&\pm \Delta_p ^* &i \left ( \partial _x +i\partial _y \right )  & \mp \mcal+\mu
\end{array} \right )  \: ,
\label{blockd}
\eneq 
 (the derivatives act on the $\psi $ fields).
 An unitary  transformation,  by changing the basis,  $\hat {B} ^T  \to  \left [  \psi_{g\uparrow}, \psi_{u\downarrow},
 \psi_{g\downarrow}  , \psi_{u\uparrow } | -\psi_{g\uparrow}\;^\dagger, \psi_{u\downarrow}\;^\dagger,
 \psi_{g\downarrow}\;^\dagger,  - \psi_{u\uparrow }\;^\dagger \right ]^T $, rewrites  this Hamiltonian into  the  BdG form
of Eq.\eqref{mat2}. It can also be shown that another unitary transformation maps the Hamiltonian of Eq.\eqref{fullpo}  into the one of Ref.\onlinecite{hsiehfu}.

A plane wave representation can be used for a translationally invariant material in $\vec{k}$ space. It is easy to check that 
$ \Xi \: H  (\vec{k} ) \:  \Xi ^T = - H  (- \vec{k} )$.  Time reversal transformation provides 
$ \Theta \: H  (\vec{k}  ) \:  \Theta ^T = H  ( -\vec{k} )$ only for  $\Delta_p  ^*= -\Delta_p $.  
  The chirality  operator,  $ \Gamma = i\:  \Xi \: \Theta  $  is such that $\Gamma H(k) \Gamma ^{-1}Ê= H(k)  $, provided   $\Delta _p^* = -\Delta_p $.  

Bulk excitation energies, $\lambda $, of the Hamiltonian of Eq.\eqref{fullpo} are given by ($k^2 =  k_\parallel ^2 + k_z^2 $):
\bea
\lambda^2 =k^2+& M^2& + \mu ^2 + |\Delta_p |^2 \label{bande}\\
&\pm & 2 \sqrt{ (  k_z^2+ M^2 )(  \mu ^2 + |\Delta_p |^2) +k_\parallel ^2 \mu ^2 }. 
\nonumber
\enea
Here we have kept only linear terms in the Hamiltonian by dropping the laplacian appearing inside  $\mcal $ (i.e.  $\mcal \to M$).  
 When the chemical potential is in the conduction band $( \mu > M >>|\Delta _p| $), we can neglect $|\Delta_p|^2  $ in the square root. Defining single particle energies  $\eta _{\vec{k}} = \sqrt{ M^2 + k^2 } $ the  eigenenergies take the form:
 \bea
 \lambda =  \pm \sqrt{\left ( \pm \eta _{\vec{k}} -\mu \right )^2 + |\Delta _p|^2 } . 
 \label{bubu}
 \enea
 While two of the resulting bands are very far from $\mu$, the other two describe the opening of the superconducting gap
 at $\eta _{\vec{k}} \sim  \mu $.   

 It is known that at an interface between a TI and a trivial insulator,  there are delocalized boundary states. Their energy dispersion is described by a Dirac cone  $ \epsilon = \pm k_\parallel$, where $k_\parallel$ lies on the boundary.  An interface between a TI and a trivial insulator with the same bulk gap, but with no band inversion, can be easily mimicked  just by keeping the linear terms in the derivatives of the model Hamiltonian of Eq.\eqref{ham0} and by changing the sign of $M$ between the two half spaces\cite{noi}. The resulting  wavefunctions  of the  boundary states are concentrated on the plane at $z=0$ and decay exponentially in the $z$ direction normal to the surface, with  the decay length  $|M|/\hbar v $. In view of the fact that the linearized model Hamiltonian just reproduces the long wavelength behavior, this boundary condition  is good enough\cite{fanzhang} . Analytically continuing the bulk eigenenergies  of Eq.\eqref{bande} with  $ k_z^2 \to -\kappa ^2 = -M^2 $, we get:
\bea
 \lambda = \pm \sqrt{ \left ( \pm k_\parallel -\mu \right )^2 + | \Delta_p |^2 } ,
 \label{enercone}
 \enea 
 which would describe surface superconductivity  with a gap opening at $|\mu| < M$.
  Again, two of the bands of the excitations  are very far off the Fermi energy, while the other two feature the opening of the superconducting gap in the Dirac dispersion of the boundary states.  
 
 \subsection*{Even-parity pairing} 
  In the chosen basis of Eq.\eqref{basic}, the uniform order parameter $\Delta_s $,  assumed to be independent of the orbital, as explained in Section II.A,  appears as an off diagonal contribution in an Hamiltonian of the following structure:
   \beq
H^s \!\!=\!\!
  \left  (  \:   \begin{array}{cc} H_+^s &   S^3\:  i\partial_z +\left  ({\cal{C}}^+ \Delta_s + h.c.\right ) \\
  S^3 \:   i\partial_z +
    \left ( {\cal{C}}^+ \Delta_s + h.c.\right )   & H_-^s 
\end{array} 
\right ),
\label{fulls}
\eneq 
where  
 \beq 
 H_{\pm}^s \!\! = \!\! \left (  \begin{array}{cccc}  \mp \mcal-\mu & i \left (\partial_x -i \partial_y \right )& 0  & 0\\
i   \left (\partial_x +i \partial_y \right ) & \pm \mcal -\mu& 0& 0 \\
 0 & 0 & \pm \mcal+\mu& i \left ( \partial _x -i\partial _y \right )  \\
 0&0 &i \left ( \partial _x +i\partial _y \right )  & \mp \mcal+\mu
\end{array} \right )  \:  .
\label{blocks}
\eneq 
 
The Hamiltonian in the absence of the vortex has the properties $ \Theta \: H(k) \Theta ^{-1}Ê= H(-k) $  and  $\Gamma H(k) \Gamma ^{-1}Ê=  H(k)  $, only if  $\Delta _s$ is real. 
Again, if the wavefunction decays as $\exp(-|M| |z|) $ at the boundary, the spectrum is given by Eq.\eqref{enercone}, with $\Delta _p \to \Delta _s $. An unitary transformation which changes the basis into
 \beq 
 UB \equiv  \left [  \psi_{g\uparrow} ,\:   \psi_{u\downarrow},  \:
\psi_{g\downarrow}^\dagger , \: - \psi_{u\uparrow }^\dagger \: | \: \psi_{u\uparrow},  \psi_{g\downarrow}, 
 \psi_{u\downarrow} ^\dagger, \;  - \psi_{g\uparrow }^\dagger\;  \right ]^T 
 \label{ubasic}
 \eneq
 transforms the Hamiltonian in the  convenient form of   Eq.s \eqref{blockd} and \eqref{fullpo}, with  $\Delta _p \to \Delta _s $. 

\section{Quasiparticle states bound at a vortex line}
As shown in the previous Section, the even-parity  and odd-parity  induced proximity give rise to the same matrix form of the model Hamiltonian, but in different bases. While the basis $B$ of Eq.\eqref{basic}  is appropriate for the odd-parity   superconducting correlations, giving rise to   Eq.s\eqref{fullpo},\eqref{blockd}, even-parity superconducting order  is described by  the same Hamiltonian matrix when  the basis is  $UB$ given by  Eq.\eqref{ubasic}.
We now  search for zero energy excitations  corresponding to quasiparticles bound to a vortex piercing an heterostructure $S/TI/S$ in the  form of a slab laying in the $ x-y $ plane, with  boundary  flat planes  at $z=0,L $.  Let the axis of the  vortex  be along the $\hat z$ axis. It is appropriate  to  move to cylindrical coordinates  with radial coordinate,  $r$, measured from the vortex singularity and azimuthal angle around the vortex axis, $\theta$.  Outside the vortex core,  the Hamiltonian now reads: 
 \beq
H \!\!=\!\! 
  \left  (  \:   \begin{array}{cc} H_+ &   S^3\: {\cal{C}}^3 \:  i\partial_z  \\
  S^3 \: {\cal{C}}^3 \:   i\partial_z   & H_-
\end{array} 
\right ),
\label{fullp}
\eneq 
where 
\begin{widetext}
 \beq 
H_{\pm} (r>\xi_o)=  
  \left (  \begin{array}{cccc}  \mp \mcal -\mu & i \:e^{-i\theta} 
\left (\partial_r -\frac{i}{r} \partial_\theta   - \frac{q}{2r}  \right )&\pm \Delta \: e^{-iq\theta }  & 0\\
 i \:e^{i\theta} 
\left (\partial_r +\frac{i}{r} \partial_\theta  + \frac{q}{2r}  \right ) & \pm \mcal-\mu & 0&\pm \Delta  \: e^{-iq\theta } \\
\pm \Delta ^* \: e^{iq\theta }  & 0 & \pm \mcal+\mu &
  -i \:e^{-i\theta} 
\left (\partial_r -\frac{i}{r} \partial_\theta + \frac{q}{2r}  \right )   \\
 0& \pm \Delta ^* \: e^{iq\theta }  & -i \:e^{i\theta} 
\left (\partial_r +\frac{i}{r} \partial_\theta - \frac{q}{2r}  \right )   & \mp \mcal+\mu 
\end{array} \right ) .
\label{trai}
\eneq
\end{widetext}
 Here  $q= \pm 1$ is the charge of the vortex and $\xi_o \sim  \hbar v /\Delta$ is the radius of the vortex core.  $\Delta $  stands for  $\Delta_s $or  $\Delta_p $, depending on the actual superconducting order.  We have implemented sign changes with respect to  Eq.s\eqref{fullpo},\eqref{blockd}, to  take care of the fact that all of  the derivatives should act  to the right  hand side. We have also added the vector potential associated to the vortex, which, far away from the vortex core,   takes the form of a pure singular gauge:
\beq
 A _r  =0,  \:\:\: A_\theta (r) =-  \frac{1}{r} \partial _\theta  \chi  \:\: ; \:\:\:   \chi = q\phi \: \frac{\theta} {2\pi}  \:\: ,
 \label{vpot}
\eneq
($\phi = hc/2e $ is the flux  unit).  The phase factor $e^{iq\theta } $ breaks the TR invariance, which  holds when $ \Delta_s $ is real ( $ \Delta_p $ is purely imaginary). The procedure to search for zero energy eigenstates of the Hamiltonian of Eq.\eqref{fullp} is sketched in Appendix A  for $s-wave$ pairing and  in Appendix  B for $odd-parity$ pairing. In the next Subsections we report the results.   
 
\subsection{$s-wave$, singlet   proximity}
  When proximity induces $ s-wave$, singlet  superconducting correlations,  an axial  vortex  of charge $q= \pm 1$, binds Majorana quasiparticles at the interface with the topologically trivial superconductors.
The  zero energy eigenstates are found along the lines sketched in Appendix A, by matching solutions  inside and outside the vortex core. Its boundary is defined as a circle of  radius $\xi_o \sim  \hbar v /\Delta$. An analytic derivation of the quantum field in the two-orbital model can be given far away from the vortex core in the limit of $\mu =0$ (midgap MBS). Two zero energy real fermion fields, localized far apart at the two boundary surfaces of the slab, $z\sim 0^+,\;L^-$, take the form, outside  the vortex core:
 \begin{widetext}
 \bea
\gamma (z \sim 0^+) && \propto    \: e^{ - \lambda \:  z} \: K_{ 1/2} \left ( \Delta_s \: r \right ) \: \: \left \{  \left [  e^{-i\theta /2}\:  e^{i \: q \theta  }Ê\:  \psi_{g\uparrow} \:  + 
   e^{i\theta /2}\:  e^{-i\: q \theta  }Ê\:   \psi_{g\uparrow}\;^\dagger  \right  ] + i\:  \left [  e^{-i \:  \theta /2 }Ê\:  e^{i\: q \: \theta }\:   \psi_{u\uparrow}  -    e^{i \:  \theta /2 }Ê\:  e^{-i\: q \: \theta }\:   \psi_{u\uparrow }\;^\dagger   \right ] \right \} , \label{majo}
\\
  \gamma (z\sim L^-) && \propto    \: e^{ - \lambda \: (L- z)} \: K_{ 1/2} \left ( \Delta_s \: r \right ) \:    \: \left \{-i\:   \left [  e^{i\theta /2}\:  e^{i \: q \theta  }Ê\:  \psi_{g\downarrow} \:  - 
   e^{-i\theta /2}\:  e^{-i\: q \theta  }Ê\:   \psi_{g\downarrow}\;^\dagger  \right  ] +   \left [  e^{i \:  \theta /2 }Ê\:  e^{i\: q \: \theta }\:   \psi_{u\downarrow}  +    e^{-i \:  \theta /2 }Ê\:  e^{-i\: q \: \theta }\:   \psi_{u\downarrow }\;^\dagger   \right ] \right \}, \nonumber \enea
\end{widetext}
  with $\lambda \sim |M| $ and $z>0$ in the TI. Here  $K_\pm $ are  the modified  Bessel functions: $ K_{ \pm 1/2} \left (wr \right) = e^{-wr } /\sqrt{wr}  $, so that the excitations  are localized also in the surface plane  and the wavefunction 
is,  of course, normalizable. The decay length scale is $w^{-1} \sim \xi_o$. 

The two Majorana excitations mix  both $u$ and $g$ orbitals and  are not eigenstates of the spin.  It is shown in Appendix A, Eq.s\eqref{mate}, that $ \gamma (z \sim 0)  $ and $\gamma (z \sim L) $  form a time reversed pair, notwithstanding the fact that the vortex breaks TR. This is a confirmation of the fact that they are neutral excitations, not influenced by the presence of the magnetic field. Having a $ e^{\pm i\theta /2} $ factor, they change sign when moved along a loop about the vortex singularity, as expected. 
  
 Inside the vortex core,  the solution requires an r-dependent order parameter $\Delta _s ( r ) $ together with  the corresponding 
  vector potential $A ( r ) $ and should be matched with the one given previously at  $ r= \xi _o$.      
   Solutions for any $\mu $ within the bulk gap,  require complex decay lengths $ \lambda = \lambda _1 + i \lambda _2 $ in the $z-$direction. One can impose zero wavefunction at $z=0$ and use  the fact that,  being the wavefunction associated with $\psi _{u\sigma}$ odd in $z$, a prefactor like the  $ e^{-\lambda_1z }\: \cos \lambda _2 z $ can appear, while, in the case of  $\psi _{g\uparrow}$, which corresponds to an even  $z-$function, the prefactor should be  of the type $ e^{-\lambda_1z }\: \sin \lambda _2 z $. 
   
   To derive MBSs at  finite $|\mu |< M  $, all the vector components should be involved and the solution can be found only numerically.   At  finite $\mu$, states extended along the vortex core can also be conceived.   They are also extended  on  the boundary surface, in the form of circular waves centered at the vortex singularity.  In fact,  in case  the solution is of the form $e^{\pm i \mu \: z }Ê$, it is easy to see that the Bessel functions solving these equations  have to be  of the first kind, i.e.  $J_{\pm1/2} \left (\Delta _s r \right ) $, or  $H^{(1,2)}_{\pm1/2} \left (\Delta _s r \right ) $, which describe  normalizable circular  states,  delocalized in $r$. This is briefly shown at the end of Appendix A. These states, which are not neutral fermionic states, could be scattering states  of the kind derived in Ref.\onlinecite{ioselevich}, to study coherent transport.  Analogous states occur when proximity is  $odd-parity$ and they will be presented at length in Appendix B for that case. The appearance of these states may be responsible for the  "vortex phase transition", which is expected to destroy the  MBS localized at each of the boundaries in $z =0,L$\cite{hosur}.

\subsection{Proximity induced odd-parity pairing }

An approach  similar  to the one of Subsection A can be used in the case of the Hamiltonian  of Eq.\eqref{fullp}, with $ odd-parity$ pairing and zero spin projection along the spin quantization axis which is pinned at the normal to the surface of the slab.  In this case, the basis is  given by Eq.\eqref{basic}.  In a film geometry a midgap solution of the kind of the one given in Eq.\eqref{majo},  localized at the interface and bound to the vortex,  can be found. However this is not a MBS.  In Appendix B we report the derivation of this zero energy eigenstate in full analogy with the $s-wave$ proximity case. One of these SABS  at $\mu =0 $ is:\begin{widetext}
  \bea
   \label{prim}
  \Psi _L \left (r>\tilde{\xi}_o, \theta ,z \right )  \propto   e^{-i \kappa z}   \: H_{  \frac{1}{2}} ^{(2)} (i\:  w r )  \cdot \: 
  \left \{ \left [  e^{i\pi /4} \:  e^{i (1+q)\theta /2 }  \psi_{u\downarrow} +  e^{-i\pi /4} \: e^{-i (1+q)\theta /2 }  
 \psi_{u\downarrow}\;^\dagger \right ]  \right .\nonumber \\
\left .  +  i\:   \left [ e^{-i\pi /4} \:   e^{-i (1-q)\theta /2 } \psi_{u\uparrow} + e^{i\pi /4} \:   e^{i (1-q)\theta /2 }  \psi_{u\uparrow }\;^\dagger  \right ] \right \} \nonumber\\
  \Psi _L \left (r<\tilde{\xi}_o, \theta ,z \right ) 
     \sim  e^{-\kappa ' z} \:    \xi (\kappa ' r )  \:  
  \left \{   \left [   e^{i (1+q)\theta /2 }   \psi_{u\downarrow} -     e^{-i (1+q)\theta /2 }  
 \psi_{u\downarrow}^\dagger \right ]   +
     \left [   e^{-i (1-q)\theta /2 } \psi_{u\uparrow} +   e^{i (1-q)\theta /2 }   \psi_{u\uparrow }^\dagger  \right ] \right \}.
 \enea 
  \end{widetext}
 Here   $z>0$ in the TI  ($M,C,c >0)$ and we have defined $\Delta _p = i \Delta ' $ with $\Delta ' $ real. We get: 
 \bea 
  -i\kappa  =-a_1+i \: a_2, \:\:\:\: iw = a_1 -i \: c \:  a_2 /C\hspace*{1.5cm} \nonumber\\
 a_1=  \sqrt{(C-c)( M+C {\Delta '}^2 ) -C^2 {\Delta '} ^2 }/(C-c), \nonumber\\
 \:\:\: a_2 =  \frac{ C \Delta '}{C-c}  , \:\:\: \kappa ' = \sqrt{\frac{M}{(C-c)}},\:\:\: \:\:\: M,C,c>0 ,\:\:  C-c >0 .\nonumber
\enea
This excitation  only involves fields referring to $u$ orbitals. 
By choosing the complementary set of  non vanishing components for the vector solution, a SABS arises, which involves   the $\psi _{g\sigma } $ fields only. 
For $r>\tilde{\xi}_o$, the function   decaying  in $z$ has also an oscillatory component. The function of $r$ is a  Hankel function $H_{1/2}^{(2)} (i \: w r ) $ of complex argument  and has  an oscillator factor $ \exp -i\:a_1 $, as well as  a decaying exponential factor $ \exp- c\: a_2 /C $.  For $r<\tilde{\xi}_o$, $ \xi ( \kappa ' r )  \sim   H^{(1)}  _{1/2}Ê (\kappa ' r ) + H^{(2)}  _{1/2}Ê (\kappa ' r ) $ is the combination of Hankel functions that   converges to zero at the origin (i.e. the point where   the order parameter vanishes).  In our "hard core" approximation, the value of $\tilde{\xi}_o$ is fixed by matching the two solutions  of Eq.\eqref{prim} at the core boundary. 

 These behaviors qualify the result  as a  SABS, which,by inspection,  is not  a MBS, but  a Dirac fermion. It is not an eigenstate of TR and it decays along the vortex core, in an oscillatory fashion.
 By using the projector $P_L = (1- \Gamma )/2 $ onto the left  $( L )$  chiral state ($ \Gamma = i \Xi \Theta $ defined in Section III), it is easy to check that the combinations given here  in the asymptotic region out of the vortex core, are $L-$chiral at $\theta =0$, i.e. of the form: $[\psi _{u\downarrow}+i \:  \psi _{u\uparrow}^\dagger]$ and $ [ \psi _{u\uparrow} - i\:  \psi _{u\downarrow}^\dagger] $.  The partner state to the one given in  Eq.\eqref{prim}, which involves   the $\psi _{g\sigma} $ field  operators, is the  $ R -$chiral mate combination  at $\theta =0 $,  i.e. :  $[ \psi _{g\uparrow} +i \: \psi _{g\downarrow}^\dagger ], [ \psi _{g\downarrow}-i \:  \psi _{g\uparrow}^\dagger] $.The spin expectation value for these quasiparticle states vanishes.  However, it is interesting to note  that just one spin component has a non vanishing  angular momentum around the vortex line. This is, in  Eq.\eqref{prim}, the down spin for $ q=1$ or the up spin  for $q=-1$, respectively. A similar feature is found in half vortex excitations of the $^3He \: A-$phase \cite{ivanov}.  
 
Away from the midgap, levels can be found  for Dirac Fermion excitations, which correspond to circular  waves propagating  at the interface inward or outward the vortex singularity and merging into the film by  travelling across the slab, along the vortex line, with a radial localization length $\hbar v/\Delta ' $.  These states involve both $u$ and $g$ orbitals. The  $U^{-1}(m=0)\:  \Psi (r,z) $  field, of spacial dependence  $\sim  e^{i\kappa z}  \: K_{\pm  \frac{1}{2}} ( \Delta ' r ) $, for $r> \xi _o$  in the topologically non trivial slab,  is derived in Appendix B [ Eq.\eqref{vig}].  The inverse length scale $\kappa \approx  \sqrt{\mu /c} $  can be found  in  Eq.\eqref{fat}. The location of these quasiparticle  levels is derived by matching the solution inside the core to the one outside it\cite{caroli,degennes}. The matching fixes the value of $\mu$ at which these excitations imply no energy cost. Similar levels split off the conduction or the valence band  and reside in the bulk gap.Their  orbital structure depends on the symmetry of the bulk band from where the quasiparticle level  splits off.

   We have checked other choices  for  the non vanishing vectorial components with no success and we conclude  that there is no possibility for a  MBS to exist, when proximity induced pairing is $odd-parity$. 
  
 \section{Summary and Discussion} 
   Topological insulators  hold the promise for future developments in low power spintronics and in quantum computing. 
 It has been argued  that  $Cu$ doped $Bi_2Se_3$ could become an odd-parity  topological superconductor when $\mu$ moves within the conduction band, because of doping\cite{fuberg}.
  In contrast, doping seems to be unable to induce superconductivity in $Bi_2Te_3$\cite{kadowaki}. Even in the case when the superconductors are conventional metals with $s-wave$ pairing, it can be questioned whether the pairing in the TI is $even-parity$ (singlet) or $odd-parity$ (triplet)  in nature.  Therefore, it is interesting to characterize  the nature of superconducting proximity at the interface $S/TI$\cite{zazunov}.
 We have considered   a  TI  slab  within the two-orbital model which has been successfully introduced to describe band inversion in various TI, particularly  $Bi_2Se_3$ and $ Bi_2 Te_3$. The two bands are made of orbitals which are even  ($ \psi _{g \sigma} $)  and odd 
 ($ \psi _{u \sigma} $) with respect to the surface plane   with the normal oriented along the z-axis , which terminates a quintuple layer at $z=0$.  The model Hamiltonian used here is applicable both to a full  topological superconducting state and to a TI with superconducting pairing correlations induced by  proximity. We perform the analytical matching of the state at the interface between the TI and a superconductor of trivial topology  essentially by  accounting only for the  the first order $z-$derivative appearing in the   linearized  Hamiltonian. Hence the matching  is performed just by imposing the continuity of the wave-function at the interface.  In the case of an  exponentially decaying wave-function on the topologically non  trivial and topologically trivial sides of the interface, this is the same as changing the sign of the mass  $M$ at the boundary. This simplified method was shown to provide results \cite{noi}  which are not inconsistent with  a full treatment\cite{fanzhang,qihuges}.  However, matching conditions at the interface $z =0$ are a minor concern in this context, as  we have dealt with both situations, $s-wave$ proximity and $odd-parity$ superconducting pairing,  on the same foot. In any other respect, our calculation includes second order  derivatives, because they influence  drastically the localization of the states in the planar dimensions parallel to the slab. The orbital shape of  the $even-parity$ order parameter 
  is  a combination  of $ \langle  \psi _{g \uparrow}  \psi _{g \downarrow}  \rangle $ and   of  $ \langle  \psi _{u \uparrow}  \psi _{u \downarrow}  \rangle $, while  the $odd-parity$  one, similar to  a "polar"  p-wave,  involves the fields  $   \psi _{g \sigma },  \psi _{u \sigma '}  $  multiplied together and  zero $z-$component of  the total spin of the pair in the odd-parity   case ("polar phase") and we expect that the spin-orbit coupling pins the $z-$spin quantization axis parallel to the  normal to the slab surface.  This feature could favor the proximity at the interface with a singlet superconductor and  makes  the comparison between the $even-parity$  singlet  and the  $odd-parity$  superconducting pairing more intriguing, in particular in the search for MBSs.  
 In both cases,  superconductivity opens up a  gap  at the chemical potential, which, in our case is immersed  in the Dirac cone dispersion of the boundary states.We have considered the case of  an axial vortex piercing the slab  along the normal to the surface  plane and we have derived  analytically  the  nature of the excitations within the superconducting gap, with a direct  comparison of  the two superconducting orderings.  Accounting for  the coordinate  $z$ in the quasiparticle wavefunctions,  has allowed us to find out whether the  bound state is localized at one of the surface boundaries, or it  is an extended state  along the vortex axis, across the slab  thickness.     
 
  Our main concern is  to give an answer to the question: when the chemical potential is  located in the bulk gap, can MBS  appear in the two-orbital TI model,  localized at the vortex singularity and squeezed at the interfaces with topologically trivial superconductors,  for any type of proximity ordering?  The answer is negative, as we briefly sum up, here below. 
  
   In the case of the  $s-wave$ superconductive proximity,  MBSs exist, localized at the vortex singularity. This fact is known  since  the work by  Fu and Kane\cite{fucane100}.  It is also known that  MBS could survive if the chemical potential penetrates the bulk bands ( $|\mu | > M $)\cite{hosur}. Here we recover the same result for $| \mu | < M $, by adopting  the two-orbital model.  There is one neutral (Majorana) fermion bound to the vortex, localized  at each interface (top/bottom)  with the superconductors of trivial topology. They decay exponentially  away from the interface and do non hybridize, if the slab is thick enough.  We give the full expression of the  fields for the simplest case that can be handled analytically, i.e. $\mu =0 $, Eq.\eqref{majo}.  They mix both orbitals   $   \psi _{g \sigma },  \psi _{u \sigma '}  $  but they do not mix  spin projections.
   It is interesting that the two partner MBS, localized at opposite interfaces, form a time reversed pair, notwithstanding the fact that TR  symmetry is broken by the vortex flux. This fact points to the neutral nature of these states.
   
    Away from $\mu =0$, all the components of the vector spinor are involved and a simple analytical solution cannot be exhibited.  However, by examining our analytical  derivation it is easy to argue that, if we move the chemical potential to non zero values,  the states turn out to be extended  Dirac fermions along the vortex axis or propagating waves  ingoing or outgoing the vortex singularity. Similar solutions have been studied in Ref.\onlinecite{ioselevich}  as the limiting case of  two MBS at opposite interfaces hybridizing  significantly,  till they eventually delocalize. 
   
 In the case   of $odd-parity$ superconductive proximity,   the vortex is unable to bind neutral excitations at the interfaces.
 To prove our conclusion,  we put  again the chemical potential at the midgap, $\mu =0$, and we follow  similar  steps as the ones that led us to confirm the presence of  MBSs in the $even-parity$ case.  States localized at the two interfaces are SABS, originating from Dirac fermions of opposite chirality. Being SABS excitations, they are localized close to the boundary surfaces and decay inside the slab  in an oscillatory way, in the region outside the vortex core (see Eq.\eqref{prim}). They are also localized  close to the vortex core, again in an oscillatory fashion. Within the vortex core, as well as in the topologically trivial superconductor, they have a simple exponential decay. The inverse localization length is $\propto |M|$, while the inverse wavelength  of the  oscillations is ruled by $\Delta_p$.  The expectation value of their spin is again zero, but, in this case,  vector components appear of different spin labels, while $g$ and $u$ orbitals are not mixed. This is the reverse of what happens in the $even-parity$ case, when   spin labels are separated, but different orbitals are mixed. It is interesting to note  that just one spin species in the vector  has a non vanishing  angular momentum around the vortex line, depending on the charge of the vortex. Being the pair a triplet spin pair,  the quasiparticle excitations can acquire features of the  $^3He$ quantum liquid.  
In Appendix B, we also derive scattering solutions away from midgap, in the form of circular waves localized at the interfaces and propagating from one interface to the other, by ingoing or outgoing the vortex singularity.  Their  orbital structure depends on the symmetry of the bulk band from where the quasiparticle level  splits off.

Having checked other possibilities, we conclude that there is no chance of having protected Majorana excitations with   $odd-parity$  superconductive proximity and zero projection of the pair spin, in the two-orbital model.

  {\it Acknowledgements:}  One of us (A.T.) acknowledges illuminating discussions  with P. Brouwer, F. Guinea and F. von Oppen. Work done with financial support from FP7/2007-2013 under the grant N. 264098 - MAMA (Multifunctioned Advanced Materials and Nanoscale Phenomena)  and   MIUR-Italy by Prin-project 2009 "Nanowire high critical temperature superconductor field-effect devices".
  
  \appendix
 \section{Zero energy modes for a  TI  slab with $ s-wave $ superconductivity induced by proximity}
The matrices $H_\pm^s$  of Eq.\eqref{trai} can be rotated  by using  the projector on angular momentum eigenvectors \cite{fukui}
 $ U_{m} (q,\theta ) =  e^{-im\theta}  \times    e^{-i\: S^3  \theta/2} \:   e^{i\:  q {\cal{C}}^3  \theta/2} $,
to get rid of the explicit $\theta $ dependence. Modes appear in  pair of states  of angular momentum $\pm m$ except for $m=0$. Therefore we must restrict our  search for Majorana modes to the subspace at $m=0$.
 We also get ${U_{0}^{-1} (q, \theta )}_{jj}  \nabla^2 _{jj}{ U_{0} (q, \theta )}_{jj}  \equiv  \left ( \partial _r + \frac{1}{2r} \right )^2 
+\partial _z^2 $, independent of $j$. Thus ( we put $\hbar v=1$ for the time being),  
 \bea  
 U_{0}^{-1} (q, \theta )  H_{\pm}  U_{0} (q, \theta ) = \hspace*{4cm}\nonumber\\
 \left (  \begin{array}{cccc}  \mp  {\cal{M}} -\mu &    i \: \partial_p &  \Delta   & 0\\
 i \:\partial_m   & \pm  {\cal{M}} -\mu & 0& 
\Delta  \\
 \Delta ^*  & 0 & \pm  {\cal{M}} +\mu &
 - i \: \partial_m   \\
 0&  \Delta ^*  & -i \:  \partial_p  & \mp  {\cal{M}} +\mu 
\end{array} \right )  .
\label{rot} 
\enea
with  derivatives $\partial _p,\partial_m $ including the vector potential, acting on the right hand side. Outside the core of the vortex, in the asymptotic region,  they coincide, as they become:
 \bea
 \partial _{p} \to  \partial_r +\frac{(1+q)}{2r} -\frac{q}{2r} ; \:  \partial_{m} \to \partial_r +\frac{(1-q)}{2r} +\frac{q}{2r}. 
 \label{part}
 \enea
It appears clearly that, well away from the vortex core, where the choice of $A_\theta $ given in Eq.\eqref{vpot} holds, the vector potential drops out of the Hamiltonian. 
 
  Let us search for a  zero energy eigenvector  of the Hamiltonian of Eq.\eqref{fullp} of the form $ f(z) \cdot \left [\xi_1 ,  \eta_2 , \xi_2' , \eta_1' | -\eta_1 ,\xi_2 ,-\eta_2' ,  \xi_1'  \right ]^T$. where $f(z) $  decays on  both sides away from   $z=0$ and $\xi,\eta$'s  are  complex functions of $r$. If $\xi _2,\xi '_2, \eta _2, \eta _2' $ are taken to be zero, the eight equations  that we get reduce to:
\bea
1: & \:\:\: -( {\cal {M}} +\mu )\:  \xi_1 -i \partial _z \eta_1 & =0, \nonumber\\
2: & \:\:\: i\partial_m \xi_1 + \Delta_s\:  \eta _1'&  =0 , \nonumber\\
3: & \:\:\: - i\partial_m \eta_1' + \Delta_s^* \xi _1&  =0 , \nonumber\\
4: & \:\:\: -( {\cal {M}}  -\mu) \: \eta_1'   + i \partial_{z} \xi_1' & = 0 ,\nonumber\\
5: & \:\:\: -( {\cal {M}}  - \mu ) \:  \eta_1 +i \partial_{z} \: \xi_1 & = 0, \nonumber\\
6: & \:\:\:-i \partial_{m}\:  \eta_1  + \Delta_s \:  \xi_1' & = 0 ,\nonumber\\
7: & \:\:\: -i   \partial_m\:  \xi_1' -\Delta_s^*\:  \eta_1 & = 0 ,\nonumber\\ 
8: & \:\:\: ( {\cal {M}} +\mu ) \:  \xi_1' + i \partial_z \: \eta_1'  & = 0. 
\label{sdinv1}
 \enea
 If $\Delta _s $ is real, a  solution can be obtained with $f(z) = e^{-\lambda z} \: (z>0$ and $\lambda >0 )$. We only report the $\mu =0 $ case which is analytically straightforward:   $ \xi _1'=  -\xi _1=-   K_{ -1/2} \left ( \Delta _s\: r \right )  $ and  $ \eta _1=  - { \eta _1'} =  - i \:  K_{ 1/2} \left ( \Delta_s \: r \right ) $, where $K_\pm $ are  the modified  Bessel functions: $ K_{ \pm 1/2} \left ( wr \right ) = e^{-wr } /\sqrt{wr}  $. The latter functions, together with  the exponentially decaying $f(z)$,
 are also eigenstates of the operator
\bea
 {\cal {M}} =  \left \{ M+ C\left ( \partial _r^2 +\frac{1}{2r} \: \partial _r -\frac{1}{4r^2}  \right ) + c\:  \partial _z^2 \right \} 
 \label{mama}
\enea
 with eigenvalue  $\lambda \sim M $,  that should be thus consistently  determined.   According to the given basis of Eq.\eqref{ubasic},  the linear combination of fields is a real fermion ($z>0$): 
\bea
  \propto     \: e^{ - \lambda\: z} \: K_{\pm 1/2} \left ( \Delta_s \: r \right ) \:  
  \cdot \: \left \{ \left [   \psi_{g\uparrow} \:  +  
  \psi_{g\uparrow}\;^\dagger  \right  ] + i \: \left [ \psi_{u\uparrow}  -  \psi_{u\uparrow }\;^\dagger   \right ] \right \} \nonumber
  \enea
   The opposite choice:  $ \xi _1= \xi _1'=\eta _1 =\eta_1'= 0 $, provides one solution decaying at the opposite side of the slab $z\sim L$, in the form:  $e^{- \lambda \:(L-z)} $. 
    Undoing the $U $ rotation and gauging away  the  vector potential, we obtain two real fermion fields. They  mix both orbitals and have vanishing expectation value of the spin projection. The spinor part is of the form, outside the vortex core:
  \begin{widetext}
\bea
\gamma (z \sim 0)  \to 
 \left . \left [ - e^{-i\theta /2}\: e^{i\pi /4}  \: e^{iq\theta } \psi_{g\uparrow},  0,0, e^{i \:  \theta /2 }Ê\:  e^{-i\pi /4}  \: e^{-iq\theta }\psi_{u\uparrow }\;^\dagger \right | 
 e^{-i \:  \theta /2 }  \:  e^{i\pi /4}\: e^{iq\theta }\psi_{u\uparrow}, 0,0,   e^{i\theta /2}\:e^{-i\pi /4} \: e^{-iq\theta }\psi_{g\uparrow}\;^\dagger  \right ] ^T  \:\:\\
  \gamma (z\sim L)
  \to  \: \left . \left [ 0, - e^{i \:  \theta /2 }Ê\: e^{-i\pi /4} \: e^{-iq\theta }   \psi_{u\downarrow} ,  e^{-i\theta /2}\: e^{i\pi /4} \: e^{iq\theta } \psi_{g\downarrow}\;^\dagger , 0\right | 0, e^{i\theta /2}\:  e^{-i\pi /4} \: e^{-iq\theta } \psi_{g\downarrow},  e^{-i \:  \theta /2 }Ê\: e^{i\pi /4} \: e^{iq\theta }  \psi_{u\downarrow }\;^\dagger, 0\right ]^T.\nonumber
  \label{mate}
  \enea
  \end{widetext}
  We have explicitly included the field operator,  from  the basis of Eq.\eqref{ubasic}, and we have put a slash in the middle of the spinor to highlight the internal structure of the states: $ \gamma (z \sim 0)  \to  \left [ \alpha |\beta \right ] $ and   $ \gamma (z \sim L)  \to  \left [ \alpha '|\beta '\right ] $. It can be seen that  the $\alpha '$ component and the  $\beta $ one are related by TR, as well as  the $\alpha $ component and the  $\beta '$ one. This is to underline that the two Majorana states form a TR pair, notwithstanding the fact that the vortex breaks TR symmetry. By adding the field components together, we get the final expressions or the MBSs reported in  Eq.s \eqref{majo}. 
  
  At finite $\mu$ a solution of Eq.s \eqref{sdinv1} is also possible, propagating along the vortex axis as $f(z) \sim e^{\pm i \mu z}$.  Again we require $  {\cal {M}}  [ \xi,\eta ] =0 $ to satisfy Eq.s \eqref{sdinv1} 1,4,5,8.  However, at difference with the  $\mu=0$ case worked out above, in which 
   $ \xi _1'=  -\xi _1 $ were real and  $ \eta _1=  - { \eta _1'}  $  were purely imaginary, all the vector components will now be real. It follows that the Bessel functions solving  Eq.s \eqref{sdinv1} 2,3,6,7 have an argument which is purely imaginary. This shows that they turn into extended waves $\sim J_{\pm 1/2} \left ( \Delta _s r \right ) $ or $\sim H_{\pm 1/2}^{(1,2)} \left ( \Delta _s r \right ) $, as mentioned at the end of  Section IV.A. 
  
  \section{Zero energy modes for   $ odd-parity $   superconductive proximity}

  {\bf \it  a)   SABSs at $\mu =0$}

 \vspace* {0.3cm} 
   
Here we tackle first  the midgap, zero energy eigenstate in close  correspondence with  the $\mu=0$ MBS  of the $s-wave$ pairing, presented in Appendix A. We will show that the solution  is a Dirac fermion, that can be qualified as a SABS. 
 Following the same procedure as for the $s-wave$ proximity, we start from the Hamiltonian  of Eq.\eqref{fullp} in cylindrical coordinates. A generic   form of the solution is:
 \begin{widetext}
 \bea
 f(z) \times  \left [\: \eta_1 \:e^{-i\chi} , \: \xi_1 e^{-i\chi '},\:- \xi_1' e^{i\chi'}, \:\eta_1' \:e^{i\chi}| -\eta_2 \:e^{i\chi},\xi_2 e^{i\chi'},\xi_2' e^{-i\chi'},  \eta_2' \:e^{-i\chi} \right ]^T.
  \label{veco}
 \enea
 The equations solved by the zero energy mode are:   
 \bea
 1:  \: & -(M+\mu ) \: \eta_1\: \: e^{-i \chi}+i \partial_{p} \xi_1 \: \:e^{-i \chi '}-\Delta \xi_1' \: \:e^{i \chi '}-i \partial _z \eta_2 e^{i \chi} & =0, \nonumber\\
2: \: & i \partial _ {m} \eta_1\: \: e^{-i \chi}+(M -\mu) \: \xi_1 \: \:e^{-i\chi '}+\Delta \eta_1'\: \: e^{i \chi}-i \partial_z \xi_2 e^{i \chi '} &  =0 , \nonumber\\
3: \: & \Delta ^* \eta_1 \: \:e^{-i \chi}-(M +\mu ) \:\xi_1'\: \: e^{i \chi '} -i\partial_{m} \eta_1'\: \: e^{i \chi}-i \partial_z \xi_2'  e^{-i \chi '}& =0 , \nonumber\\
4: \: & \Delta^*  \xi _1\: \:e^{-i \chi '}+ i \partial_{p} \xi_1 '\: \: e^{i \chi '}-(M -\mu) \: \eta_1'\: \: e^{i \chi}+i \partial_z \eta_2' e^{-i \chi} & = 0 , \nonumber\\
 5:  \: & i \partial_z \eta _1\: \:e^{-i \chi}-(M - \mu ) \:  \eta_2 e^{i \chi}+i \partial_{p} \xi_2 e^{i \chi '}-\Delta \xi_2' e^{-i \chi '}& =0 ,  \nonumber\\
  6:  \: & -i \partial_z \xi_1\: \: e^{-i \chi '}-i \partial_{m} \eta_2 e^{i \chi}-(M+\mu ) \:  \xi_2 e^{i \chi '}-\Delta \eta_2' e^{-i \chi} & =0 , \nonumber\\ 
  7:  \: & i \partial_z \xi_1' \: \: e^{i \chi '}+\Delta^* \eta_2 e^{i \chi}-(M-\mu )  \xi_2' e^{-i \chi '}-i \partial_{m} \eta_2' e^{-i \chi} & =0 , \nonumber\\
  8:  \: & i \partial_z \eta_1'\: \: e^{i \chi}-\Delta^* \xi_2 e^{i \chi '}-i \partial_{p} \xi_2' e^{-i \chi '}+(M+\mu )\: \eta_2' e^{-i \chi}& = 0 \:\: . 
  \label{trov3}
\enea
\end{widetext}
 We search for a real eigenvector. As 
 $\Delta_p  \sim  \langle \psi _{g\uparrow}  \psi _{u\downarrow} \rangle $  the choice $\chi = \chi ' $ is consistent with the $odd-parity$ pairing, as it gives a purely imaginary $\Delta_p \equiv i \: \Delta ' $ ( which defines $\Delta' $ as real).  By posing $f(z) = e^{-i\kappa z } $ ( with complex $\kappa$), the appropriate "would be MBS "  requires that,  e.g.,  $ \eta_1 =  \eta_1' =  \xi_2= \xi _2' = 0  $.  In this case, from Eq.s\eqref{trov3} we get: 
     \bea
 2: \:\: ( {\cal{M}} - \mu)  \xi_1  = 0   \:\:\:\:\:\:  &  5:  \:\: - ( {\cal{M}} -\mu)  \eta_2  = 0\nonumber\\
  3: \:\: -( {\cal{M}} +\mu)  \xi_1'     = 0   \:\:\:\:\:\: &  8:  \:\: ( {\cal{M}} +\mu)  \eta_2'  = 0 \label{moi},
\label{dinv}
 \enea
which implies  that, at $\mu =0$,  the surviving vector components should be eigenstates of the operator $  {\cal{M}} $ with zero eigenvalue. Besides:
  \bea
1:& \:\: i \partial _{p} \xi_1  -i \:\Delta '\: \xi_1' - \kappa \: \eta _2  & = 0 \nonumber\\
4:& \:\: i \: \partial _{p} \xi_1' -i \: \Delta '\: \xi_1+  \kappa \:  \eta _2'  &  = 0 \nonumber\\
 6:& \:\:-i\: \partial _{m} \eta_2 -i\: \Delta ' \: \eta _2' -  \kappa \: \xi_1  & = 0 \nonumber\\
 7:& \:\: -i \:\partial _{m} \eta_2'  -i\: \Delta '\: \eta_2  +\kappa \: \xi_1 ' & = 0
\label{dinv4}
 \enea
 A possible choice is  $  \xi _1= i \: \xi $,  $  \: \eta_2 = i\: \eta  $ and $  \xi _1'= \eta $,  $  \: \eta_2' =\xi   $. This implies that the equations become:
   \bea
   1: & \:\:  \: \partial _{p} \xi + i \: (  \kappa +  \Delta ' ) \:  \eta    = 0, \:\:\: \nonumber\\
4: & \:\:  \: \partial _{p} \eta -i \: (   \kappa + \Delta ' )\: \xi   = 0,\:\:\:  \nonumber\\
 6: & \: \:\:\partial _{m} \eta -i\: ( \kappa  + \Delta ' ) \: \xi   = 0 \nonumber\\
 7:  & \: \:\: \partial _{m} \xi +i \: ( \kappa +\Delta ') \:  \eta   = 0.
\label{dinve}
 \enea
Rewriting  $ \partial = \partial _r + 1/2r $
 in place of  $\partial _{m,p} $, because, asymptotically out of the vortex core,  the two operators  coincide,  a good solution out of  the vortex core, satisfying convergency for large $r$  is:  $\eta  \sim H_{-1/2}^{(1)} (-i\: wr) $  and $\xi  \sim H_{1/2}^{(1)}  (-i\: wr) $ 
with    $w = (\kappa + \Delta ' ) $.  The latter turns out to be a complex parameter. Observing that $ \left (\partial - \alpha ^2 \right ) \: H_{\pm 1/2}^{(1)} \left ( \alpha   r \right )=0 $, the constraint  $  {\cal{M}}  \: H_{\pm 1/2}^{(1)}=0 $ is fulfilled, provided $
  M +C \: w^2  -c \: \kappa ^2=0 $, what gives ($ a_1 =( a+a^*)/2, a_2= (a-a^*)/2 $):
  \bea 
  -i\kappa  =-a_1+i \: a_2, \:\:\:\: -iw = -a_1 +i \: c \:  a_2 /C \nonumber\\
 a_1=  \sqrt{ ( C-c) ( M+C {\Delta '}^2 ) -C^2 {\Delta '} ^2 } /(C-c)  , \nonumber\\
 \:\:\: a_2 =  \frac{ C \Delta '}{C-c}  , \:\:\: M,C,c >0 ,\:\:  C-c >0 .\nonumber
\enea
  
Accordingly, $f(z)$ is a decaying  function  of  $z$, but it  also has an oscillatory component. The function of $r$ can be put in the form of the  Hankel function $H_{1/2}^{(2)} (i \: w r ) $ and has  an oscillator factor $ \exp -i\:a_1 $, as well as  a decaying exponential $ \exp- c\: a_2 /C $. This behavior qualifies the state to be a SABS, which is not  a MBS, however. In fact, being  the wavefunction  of the form $ \left [\: 0,i\:  \xi ,\xi , 0| i\: \xi , 0, 0,  -  \xi  \right ]$,
in the basis of Eq.\eqref{basic} the field operator of the excitation is:
\bea
 e^{-i\kappa z}  \: H_{  \frac{1}{2}} ^{(2)} (i\:  w r )  \cdot \: \left \{
    \left [e^{i\pi /4} \: \psi_{u\downarrow}+e^{-i\pi /4} \: 
 \psi_{u\downarrow}\;^\dagger\right ]  \right . \nonumber\\
 \left . +i \: \left [ e^{-i\pi /4} \: \psi_{u\uparrow}+ e^{i\pi /4} \:   \psi_{u\uparrow }\;^\dagger \right ] \right \}.
 \label{truc}
 \enea
 which correspond to  a mid gap SABS. 
  
 Inside the vortex core we  do not require an exponential decay in $r$, but a zero of the wavefunction  at $r=0$.   Let us define 
 $f(z) =  e^{- \kappa' z }$ for $z>0$. In this case  the Eq.s\eqref{trov3} become: 
  \bea
1:& \:\:\left [  i \partial _{p} \xi_1  -i \:\Delta '\: \xi_1'  + i \kappa '\: \eta _2 \right ]   & = 0 \nonumber\\
4:& \:\: \left [ i \: \partial _{p} \xi_1' -i \: \Delta '\: \xi_1  - i \kappa '\:  \eta _2' \right ]  &  = 0 \nonumber\\
 6:& \:\:\left [-i\: \partial _{m} \eta_2 -i\: \Delta ' \: \eta _2' + i\:  \kappa '\: \xi_1 \right ]  & = 0 \nonumber\\
 7:& \:\: \left [-i \:\partial _{m} \eta_2'  -i\: \Delta '\: \eta_2  -i \: \kappa '\: \xi_1 '   \right ] & = 0.
\label{dinv5}
 \enea
 Here $\partial _{p,m}$ do not have the form of Eq.\eqref{part}, because,  the vector potential does not take the asymptotic expression typical of a singular gauge within the vortex core. Also  $\Delta '  ( r )$ is expected to have a linear $r-$ dependence close to the vortex axis\cite{degennes}.
 A simple way to approximate the eigenfunction is to consider a hard core with   $\Delta '=0 $ and to overlook the vector potential difference. By putting $ \eta_2' = \xi_1 \equiv \xi $ and $  \xi _1' =   \eta_2\equiv \eta  $  we get  the two equations: 
   \bea
1  (\sim 7):  & \:\:\:   \partial   \:  \xi  +   \kappa '   \:  \eta= 0   \hspace*{3cm} \nonumber\\
 4 (  \sim 6):   & \:\:\:     \partial \:   \eta  -  \kappa '  Ê \: \xi=0 \hspace*{3cm} \nonumber\\
\to   &  \partial ^2  \xi  =  -{\kappa '}^2 \:\xi  , \:\: \:    \partial ^2  \eta =  -{\kappa '}^2 \:\eta  . 
   \label{fcorp4}
 \enea
 Appropriate solutions of Eqs.\eqref{fcorp4}  are the Hankel functions  $H^{(1,2)}  _{\pm 1/2}Ê (\kappa ' r ) $.
 Besides
\beq
 {\cal{M}} \: \left [ \xi ,\eta \right]  =0  \: \to M - (C -c)\: {\kappa '}^2  =0,
 \eneq 
 what defines $\kappa ' $ as a real parameter for $M,C-c >0 $.
 Having two possible normalizable solutions, we can  impose that their combination vanishes at $r=0$. 
We thus  obtain $ \xi(r) $ and $\eta (r) $:
 \bea
\xi(r)  =  H^{(1)}  _{+ 1/2}Ê (\alpha r ) + H^{(2)}  _{+ 1/2}Ê (\alpha r )  \: , 
\nonumber\\
 \eta (r)  = - i \left \{   H^{(1)} _{- 1/2}Ê (\alpha r ) -  H^{(2)} _{- 1/2}Ê (\alpha' r ) \right \},
\enea 
which are  real for any  $r$. They  converge to zero at the origin and are such that $ \eta (r ) =  \xi (r) $.  
In conclusion,  inside  the   core, the vector is:  
   \bea
    \sim    e^ {-\kappa'  z}\:  \left .    \left [  0,  \xi(r ) , -   \eta (r )   ,0 \right  |  -  \eta (r )  , 0, 0,   \xi (r )  \right ]^T.
   \enea
  In the basis of Eq.\eqref{basic}, this gives:
  \bea
 \sim   \xi (r) \:    e^ {-\kappa'  z}\:    \left \{    \left ( \:    \psi_{u\downarrow} -
 \psi_{u\downarrow}\;^\dagger \right ) -    \left ( \psi_{u\uparrow} +   \psi_{u\uparrow }\;^\dagger  \right ) \right \} .
  \enea 
  Undoing the transformation $U(m=0)$ we get Eq.\eqref{prim} of the text. 
  Our   "hard core" approximation depends on the parameter $\tilde{\xi}_o$, which fixes the core boundary.  The value of $\tilde{\xi}_o$ can be  determined by matching the inside and the outside  solutions at the core boundary. 
  
  By choosing $ \eta_1 ,  \eta_1' ,  \xi_2, \xi _2'  $ non zero and the other spinorial components zero,  the partner state of Eq.\eqref{prim}, involving the  $\psi _{g\sigma } $ fields, can be found.    
 
 \vspace* {0.3cm} 

  {\bf  \it  b) extended states  along the vortex line}
 
  \vspace* {0.3cm} 
 
 Here  we show that a  Fermi Dirac state solution is also possible, involving real vector components, which describe a wave   travelling along the vortex , with  $z-$dependence $e^{i\kappa z } $ and $\kappa $ real.   This is a bound state within the bulk gap.  All the $\xi 's ,\eta's $ have to be non vanishing. If we pose $ \eta_1'=- \eta _2, \:  \eta_2'= -\eta_1 $ and  $ \xi _1' = \xi _2 , \: \xi_2' = \xi _1 $, the Eq.\eqref{trov3} can be solved  with:
 \bea
 \xi _1 = K_{1/2} ( \Delta ' r ),\:\:\:\: \xi _2 = K_{-1/2} ( \Delta ' r )  \nonumber\\
  \eta _1 = K_{1/2} ( \Delta ' r ),\:\:\:\:   \eta _2 = K_{-1/2} ( \Delta ' r ) 
 \enea
by  requiring the following :
 \bea
 1: \:\: -({\cal{M}}+\mu)  \eta_1  + \kappa \: \eta_2   = 0   \:\: &  5:  \:\: -({\cal{M}}-\mu)  \eta_2 - \kappa \: \eta_1  = 0\nonumber\\
 provided \nonumber\\
 1': \:\:  \partial _{p} \xi_1  -\Delta' \: \xi_1'  = 0  \:\:\:\:\:\:  & 5':  \:\:  \partial _{p} \xi_2  -\Delta '\: \xi_2'  = 0  
\nonumber 
\enea
The first two are compatible, if, after substitution of the  eigenvalue  to the operator $ {\cal{M}} $, the determinant : $ {\cal{M}}^2 -\mu ^2  + \kappa ^2 =0 $  vanishes.
Similarly: 
\bea
8: \:\: -({\cal{M}}+\mu)  \eta_2'  + \kappa \: \eta_1'   = 0   \:\: &  4:  \:\: -({\cal{M}}-\mu)  \eta_1' - \kappa \: \eta_2'  = 0\nonumber\\
 provided\nonumber\\
 8': \:\:  \partial _{p} \xi_2'  -\Delta' \: \xi_2  = 0  \:\: & 4':  \:\:  \partial _{p} \xi_1'  -\Delta '\: \xi_1  = 0  
\nonumber\\
6: \:\: -({\cal{M}}+\mu)  \xi_2  + \kappa \: \xi_1   = 0   \:\: &  2:  \:\: ({\cal{M}}-\mu)  \xi_1 + \kappa \: \xi_2  = 0\nonumber\\
 provided\nonumber\\
 6':  \:\: -i \partial _{m} \eta_2  +\Delta '\: \eta_2'  = 0  \:\:&  2': \:\:  \partial _{m} \eta_1 +\Delta' \: \eta_1'    = 0 
\nonumber\\
 3: \:\: -({\cal{M}}+\mu)  \xi_1'  + \kappa \: \xi_2'   = 0   \:\:  &  7:  \:\: ({\cal{M}}-\mu)  \xi_2' + \kappa \: \xi_1'  = 0\nonumber\\
 provided\nonumber\\
 3':  \:\:  \partial _{m} \eta_1'  +\Delta '\: \eta_1 = 0  \:\:&  7': \:\:  \partial _{m} \eta_2' +\Delta' \: \eta_2   = 0 
\nonumber
\enea
The   value of $\kappa $ for any $\mu$ is fixed by solving the determinantal condition: 
\beq
\left [M + C {\Delta ' }^2 -  c \kappa ^2 \right]^2 +\left ( \hbar v\: \kappa\right )  ^2 -\mu ^2 = 0
\eneq
( $\hbar v$ has been restored). By posing $ m= M + C{ \Delta ' }^2 $  we get:
\bea
 \kappa^2  =  \frac{ m}{c} -\frac{\left ( \hbar v\right )  ^2}{2c^2}  +  \frac{1}{c}  \sqrt{ \mu ^2 - \frac{m\left ( \hbar v\right )  ^2}{c} +\frac{\left ( \hbar v\right )^4}{4c^2} }  
 \label{fat}
\enea
which gives a real $\kappa $, because\cite{nota2}:
 $ m >  \left ( \hbar v\right )  ^2/ 2c $. 
 The plus sign in front of the square root of Eq.\eqref{fat} has been chosen on physical grounds,  in order to obtain that the wavelength of the propagation along $z$ increases when  $\mu $ goes  deeper down in the bulk gap. Qualitatively is $\kappa \sim \sqrt{ \mu /c } $. The matching between the inside and the outside of the vortex core will fix the value of $\mu$ at which the excitation is zero energy\cite{caroli}. Hence the location of the energy level is fully determined. 
 
Just outside the slab, at the interface with  the topologically trivial material is  $m<0 $, so that $  \kappa^2  <0 $ and the solution decays with $z$ away from the surface. Being $\kappa $ purely imaginary,   the vector components of the eigenfunction acquires alternatively an $i$ factor, as is the case leading to Eq.\eqref{truc}. This changes the Bessel functions $K_{\pm 1/2}$ into the corresponding Hankel functions. While in the case of $\mu=0$ these functions were localized in $r$ anyhow, because of complex argument, in this case they  are delocalized in $r$, because the argument, $\Delta 'r $, is real. This implies that, at the interface, the waves become circularly propagating inward or outward the vortex line. 
  
 The consequence is that this state   is a traveling wave along the vortex axis in the non trivial topological material , while it decays outside at the interface with  the trivial material and propagates outward  or inward at the surface.  Its energy is  localized  in the gap.  
 This is not a Majorana state, however.  Using  Eq.\eqref{veco} for the vector, we obtain: 
  \begin{widetext}
 \bea
  \left [\: \eta _1, \xi _1,- \xi _2 , -\eta _2| -\eta _2, \xi _2, \xi _1, - \eta _1  \right ]    \to
    \left [\: \eta , \xi ,- \xi  , -\eta | -\eta , \xi , \xi , - \eta   \right ] \nonumber 
 \enea
which, in the basis of Eq.\eqref{basic},  provides:
\bea
  \to  U^{-1}(m=0)\: \Psi ( r,z) \propto e^{i\kappa z}  \: K_{\pm  \frac{1}{2}} ( \Delta ' r )  \cdot \: 
    \left [\left ( \psi_{g\uparrow}+
 \psi_{g\uparrow}\;^\dagger\right )+  \left ( \psi_{u\downarrow}-
 \psi_{u\downarrow}\;^\dagger\right )+ \left ( \psi_{u\uparrow}+
 \psi_{u\uparrow}\;^\dagger\right )  + \left ( \psi_{g\downarrow}+
 \psi_{g\downarrow}\;^\dagger\right )
 \right ] .
 \label{vig}
 \enea
 The final step would be undoing the transformation $U(m=0)$.   
 
 \end{widetext}


\begin{thebibliography}{99}
\bibitem{not}  this is unitary equivalent to the BdG form of Eq.\eqref{mat2}  given below.
\bibitem{nota2}  The parameters   for $Bi_2 Se_3$, have been estimated as $ M= 0.28 eV $,$C = 56.6 \: eV \AA ^2 $, $c = 10 \: eV \AA ^2 $, $v= 5\times 10^5 m/sec $, $\Delta  \sim 3.5 ¡K \sim 0.25 meV$\cite{zhang_nphys}.
\end{thebibliography}
\end{document}